%% file: Inductorreview.tex
\documentclass[journal,twoside]{IEEEtran}
\usepackage{cite}

\ifCLASSINFOpdf
\else
\fi
\ifCLASSOPTIONcompsoc
  \usepackage[caption=false,font=normalsize,labelfont=sf,textfont=sf]{subfig}
\else
  \usepackage[caption=false,font=footnotesize]{subfig}
\usepackage[mediumspace]{SIunits}
\usepackage{circuitikz}
\usepackage{multirow} 
\usepackage{rotating} 
\usepackage{colortbl} 
\usepackage[nonumberlist,toc,nopostdot,acronym,symbols,nomain,hyperfirst=false]{glossaries}

\loadglsentries{PPgloss}

\begin{document}
\bstctlcite{IEEEexample:BSTcontrol}
%
\title{Review of Inductive Pulsed Power Generators for Railguns}

%

\author{Oliver~Liebfried
\thanks{Manuscript received Jan. 15, 2017; accepted Mar. 20, 2017. Date of current version Mar. 21, 2017.}%
\thanks{O. Liebfried is with the French-German Research Institute of Saint-Louis (ISL), 5 rue G\'en\'eral Cassagnou, 68301 Saint-Louis, France, e-mail: Oliver.Liebfried@isl.eu.}
\thanks{The published version of the paper will be available online at http://ieeexplore.ieee.org.}%
\thanks{Digital Object Identifier 10.1109/TPS.2017.2686648}}

%
%
%

\markboth{Accepted by IEEE Transactions on Plasma Science,
~March~2017}%
{Liebfried: Review of Inductive Pulsed Power Generators for Railguns}
%

\IEEEpubid{
\copyright~2017 IEEE}

\pagestyle{empty}
\vspace{5cm}
\begin{table*}
	\centering
		\begin{tabular}{c}
This work  has been accepted for publication by IEEE Transactions on Plasma Science. \\The published version of the paper will be available online at http://ieeexplore.ieee.org. It can be accessed by using the following \\Digital Object Identifier: 10.1109/TPS.2017.2686648.\\\\\\

\textcopyright  2017 IEEE. Personal use of this material is permitted. Permission from IEEE must be obtained for all other\\
uses, including reprinting/republishing this material for advertising or promotional purposes, collecting new\\
collected works for resale or redistribution to servers or lists, or reuse of any copyrighted component of\\
this work in other works.
		\end{tabular}
\end{table*}
\clearpage
\setcounter{page}{1}

\maketitle

\begin{abstract}
This literature review addresses inductive pulsed power generators and their major components. Different inductive storage designs like solenoids, toroids and force-balanced coils are briefly presented and their advantages and disadvantages are mentioned. Special emphasis is given to inductive circuit topologies which have been investigated in railgun research such as the XRAM, meat grinder or pulse transformer topologies. One section deals with opening switches as they are indispensable for inductive storages and another one deals briefly with SMES for pulsed power applications. In the end, the most relevant inductor systems which were realized in respect to railgun research are summarized in a table, together with its main characteristics.

\end{abstract}


%
\IEEEpeerreviewmaketitle

\section{Introduction}
\label{ch:Introduction}
%
%
%
%
%
\IEEEPARstart{I}{nductive} storages are used in pulsed power generators for railguns for several reasons. In combination with a \gls{HPG} or a battery, inductors are used to generate the high voltage which is needed to inject current into a railgun at high armature velocities. They are used in capacitor banks to limit the maximum current amplitude and adjust the pulse length to the requirements of the railgun. Furthermore, they decouple several capacitor modules from each other and allow time-delayed switching and therefore pulse shaping. However, inductor research was not very evident within the electromagnetic launch community in the past as inductors were mostly part of HPG, capacitor bank, or battery system development. Thus, there is no comprehensive reference work about inductive storage systems for pulsed power generation. This paper aims at filling the gap as inductive pulsed power systems are becoming more relevant due to emerging technologies like Li-ion batteries, supercapacitors, and superconductors.

\section{Inductive storage}
Charged inductors can be seen as current sources which can create any voltage, assuming a corresponding insulation. Thus, inductors represent an ideal power source for railguns. In an inductive storage, energy is stored by its magnetic field. The interaction between the magnetic field and the current in the windings creates Lorentz forces on the windings of the inductor. Therefore, a coil can be regarded as a pressure vessel with the magnetic field $B$ as a pressurized medium. The corresponding pressure $p$ is related by $p=\frac{1}{2 \mu} B^2$ to the magnetic field $B$ with the permeability $\mu$. 
The energy density of the inductor is directly linked to the magnetic field and therefore, its maximum depends on the tensile strength of the windings and the mechanical support. The upper limit for the energy density $W/V$ follows from the virial theorem and can be expressed by
\begin{equation}
	W/V= \lambda \sigma
\end{equation}
where $\sigma$ is the tensile strength of the coil material and $\lambda$ is a coefficient related to the structural efficiency of the coil design \cite{Gredin1985,Nomura2005}. $\lambda=1$ is the maximum value which is achievable by an ideal coil geometry. In the case of long and thin solenoidal coils or toroidal coils, $\lambda$ gets close to $\frac{1}{3}$ \cite{Moon1982,Eyssa1981}. In practice, the best value for $\lambda$ can be achieved with short solenoidal coils like the Brooks coil, where $\lambda$ can be as high as $\frac{1}{2}$ \cite{Eyssa1981,Gredin1985}. 

In superconducting coils, the mechanical stress is the major limitation of the energy density. In normal conducting coils, other limitations like resistive losses or heat dissipation might be more critical than the mechanical stress.


\IEEEpubidadjcol
\section{Coil design}

\begin{figure*}[!t]
\centering
\subfloat[Brooks coil (cutted view)]{
\begin{tikzpicture}[scale=1]%
  \footnotesize%
    \draw (0mm,0mm) node[anchor=south west]{\includegraphics[width=1.5in]{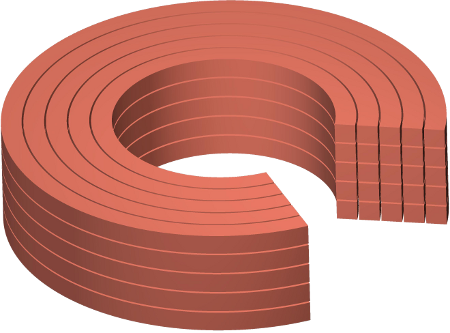}};%
    \draw[stealth-stealth, thick] (2.cm,2.06cm) -- (2.94cm,2.00cm);%
    \draw[stealth-stealth, thick] (2.94cm,2cm) -- (3.87cm,1.95cm);%
    \draw (3.87cm,1.83cm) -- (3.87cm,2.05cm);%
    \draw (2.94cm,1.89cm) -- (2.94cm,2.1cm);%
    \draw (3.9cm,1.81cm) -- (4.1cm,1.8cm);%
    \draw (3.9cm,1.02cm) -- (4.1cm,1.01cm);%
    \draw [stealth-stealth, thick] (4cm,1.01cm) -- (4cm,1.8cm);%
    \draw [dash pattern=on 1pt off 2pt on 5pt off 2pt,thick] (2.cm,1.45cm) -- (2cm,3cm);%
    \draw [dash pattern=on 1pt off 2pt on 5pt off 2pt,thick] (2cm,0.1cm) -- (2cm,-0cm);%
    \draw (3.93cm,1.43cm) node[anchor=west] {$a$};%
    \draw (2.5cm,2.12cm) node {$a$};%
    \draw (3.4cm,2.07cm) node {$a$};%
  \end{tikzpicture}%

\label{fig:Brooks}}
\hfill
\subfloat[Concentric solenoids]{\includegraphics[width=0.7in]{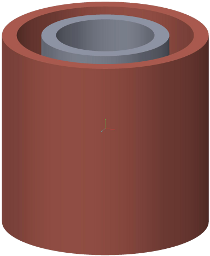}%
\label{fig:concentric}}
\hfill
\subfloat[Hexagonal solenoids]{\includegraphics[width=1.2in]{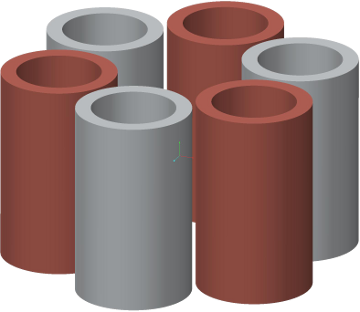}%
\label{fig:hex}}
\hfill
\subfloat[D-shape toroid]{\includegraphics[width=1.2in]{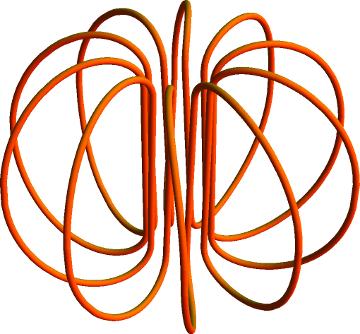}%
\label{fig:D-shape}}
\hfill
\subfloat[Force-balanced coil]{\includegraphics[width=1.5in]{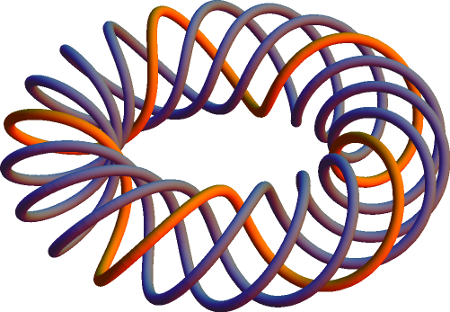}%
\label{fig:FBC}}
\caption{Different coil geometries and arrangements}
\label{fig:1}
\end{figure*}
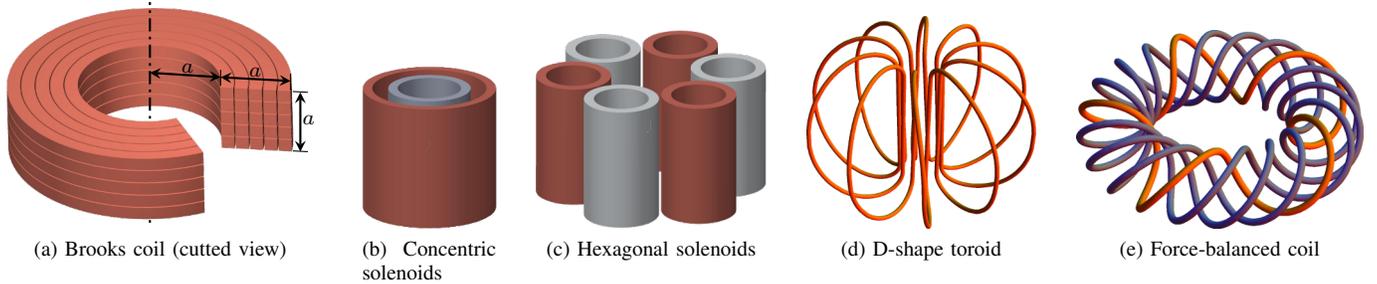 

The optimization of normal conducting coils with respect to the energy density and $L/R$ is usually performed by optimizing the inductance per \gls{GPW} \cite{Murgatroyd1989}. The coil with the best inductance per \gls{GPW} is known as the {\bfseries Brooks coil} (see Fig.~\ref{fig:Brooks}) which is a solenoidal coil where the inner radius corresponds to half the outer radius and is equal to the height of the coil 
\cite{Murgatroyd1989,Grover1946}. Additionally, the number of windings in the radial and axial directions is the same.

A significant drawback of solenoids is that very high fringe fields appear in the vicinity of the coil. A safety distance between the coil and the operator or sensitive electronics is necessary in most cases if magnetic shielding is not applied \cite{Polk1992}. But even then, a certain distance between the coil and the shielding has to be kept in order not to saturate the shielding material \cite{Pokryvailo2003}. Placing several solenoids in a concentric or hexagonal arrangement (see Fig.~\ref{fig:hex} \& \ref{fig:concentric}) can help to reduce fringe fields \cite{Weck1997,Schoenwetter1995}.

An alternative to the solenoidal coil is the {\bfseries toroidal coil}. Here, the magnetic field is encased in the windings of the coil and, therefore, a fringe field, ideally, does not exist. In actual fact, a fringe field cannot be completely avoided but a toroid made of five or more solenoids has a fringe field at least ten times smaller than a solenoidal coil \cite{Kanter1995}. Thus, shielding or a safety distance might not be necessary. 
On the other hand, the encapsulation of the magnetic field is attained at the expense of the energy density. The inductance per \gls{GPW} of a solenoidal filament coil is, for example, 2.6 times higher than that of a toroidal filament coil with a circular turn window and a totally filled coil center (toroidal counterpart of a Brooks coil) \cite{Murgatroyd1989}. The inductance of toroids can be further increased by choosing the so-called {\itshape Shafranov D}- or {\itshape Princeton D}-shape for the windings (see Fig.~\ref{fig:D-shape}) \cite{Murgatroyd1989,shafranov1973,birkner1991}. The D-shape is not only the optimum shape with respect to the energy density but also an optimum with respect to the mechanical stress distribution.

Recent optimizations with respect to the mechanical stresses performed in superconducting coils have led to the development of the {\bfseries force-balanced coil} (see Fig.~\ref{fig:FBC}). It is a combination of a solenoidal coil with a toroidal one realized by adapting an optimized pitch angle to the helical windings of a toroidal coil \cite{Nomura2005}. It offers a good compromise between storage efficiency and electromagnetic compatibility. A drawback is that producing it is very complex.

\section{Opening Switches}

In order to transfer the energy from an inductive storage inductor to a load, an opening switch is necessary which interrupts the charging circuit and commutates the current to the load. During the commutation process, a voltage $V$ depending on the self-inductance $L$ of the circuit and the load current $I$ is generated equally at the load and the opening switch according to $V=L\frac{dI}{dt}$. This voltage generation is desired when batteries or \glspl{HPG} are used which are generally low voltage sources. But the same voltage is applied to the opening switch and in case of large railguns, the so called speed-voltage can easily amount to several kilovolts \cite[p.~8]{Marshall2004}. Therefore, opening switches for high current applications are challenging and were often avoided by preferring capacitor based solutions with closing switches.

Three kinds of opening switches are basically available: mechanical switches, fuses, turn-off controlled semiconducting switches and hybrids of all of them \cite{Pokryvailo1999,Schoenbach1984,Bluhm2006}. For inductively powered railgun systems, heavy duty contactors \cite{Sterrett1987}, explosively driven switches \cite{Aivaliotis1989,Peterson1991,Klug1991} and counter-current switch-off techniques \cite{VanDijk1995,Leung1989,Ding2012,Fridman2013} were used. Today, considerations go toward semiconducting switches like \acrshortpl{GTO} or \acrshortpl{IGBT} which are now capable to operate for short time in the MW range \cite{Scharnholz2003b,Sitzman2011}. The recent availability of such improved components facilitates system designs, but in order to switch off currents in the kA range, arrays of such switches are still necessary, which is expensive, heavy and bulky \cite{Liebfried2013c}. Wide bandgap semiconductor switches based on materials like \acrshort{SiC}, \acrshort{GaN} or diamond promise better electrical properties but the state of the art is still far from required values. An applicable technique in an inductive pulsed power generator is the introduction of an inverse current to a thyristor, allowing the switch to turn-off safely at a low or zero current. Efforts at ISL in this direction resulted in the development of the so-called \acrshort{ICCOS} switch which utilizes power thyristors in this respect \cite{Dedie2009b,Brommer2009Patent}. Such switch is avoiding the generation of high commutation voltages by applying the inverse current also to parasitic circuit inductances. A design study for a 1\,MJ inductive generator clearly showed the advantages of this switch compared to other semiconductor switches \cite{Liebfried2013c} with respect to size, weight and costs. Researchers at \gls{TNO} built and tested such kind of opening switch for currents up to 300\,kA for their \gls{HPG} system \cite{vanDijk1997}.


\section{Pulse generation circuits used for railguns}
\subsection{Single coil}
Single coil setups have been built in combination with HPGs \cite{Gully1983,Hackworth1986} or huge battery banks \cite{Sterrett1987,Pokryvailo1998,Pokryvailo2005}. They are needed for the generation of high voltages, as HPGs and batteries are usually operated at a few hundred volts. 
Inductances are also widely used in capacitive pulsed-power generators \cite{Spahn1995a,Wolfe2004}. But in this case, the aim is to achieve pulse shaping in order to generate the required pulse length and to limit the maximum current. In contrast to other prime energy sources, opening switches are not required.

\begin{figure}
  \footnotesize
  \centering
  \input{Figs/Inductive/Singlecoilcircuit.tex}
  \caption{Simplyfied circuit for a single coil pulsed power generator}
  \label{fig:singlecoil}
\end{figure}
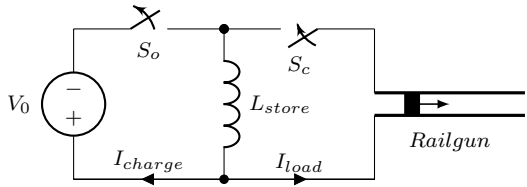

Fig.~\ref{fig:singlecoil} displays a circuit diagram which shows how a railgun can be supplied by a single coil. First, the inductor $L_{store}$ is charged by the voltage source $V_0$. After charging, the current is commutated to the railgun by opening switch $S_o$ and closing switch $S_c$ at the same time.

\ctikzset{bipoles/length=1.4cm}

In single coil setups, the amplitude of the load current directly correlates with the maximum current with which the inductance is charged. Current amplifying circuit topologies could be a better option, as they need a less powerful current source and opening switch. The following three basic circuit topologies are repeatedly suggested in the field of railgun applications.

\subsection{Pulse transformer}

\begin{figure}[tb]
  \footnotesize
  \centering
  \input{Figs/Inductive/Transformercircuit.tex}
  \caption{Simplyfied pulse transformer circuit}
  \label{fig:pulsetransformer}
\end{figure}
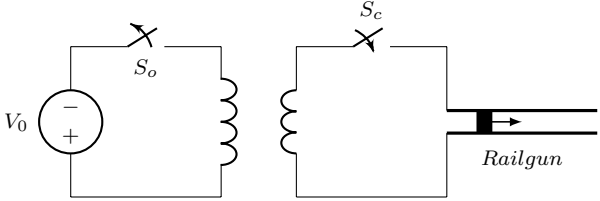

Pulse transformers are based on the magnetic energy transfer between two coupled coils. Transformers are usually used for voltage amplification, but if the turns ratio between the secondary and primary windings is less than 1 ($N_2/N_1<1$), they can also be used for current amplification. But note that a transformer itself is not a power amplificator because the transformation ratios of current and voltage are invers to each other. Pulses for railguns can be generated by
discharging capacitors into the primary windings of the transformer \cite{Woffort1991,Day1993,Wu2013c,Li2012b} or by applying pulse compression if a DC voltage source is used for charging the primary windings of the transformer as shown in Fig.~\ref{fig:pulsetransformer}. Pulse compression can be achieved by using an opening switch $S_o$ to decrease the current in the primary windings rapidly, thus generating a high current in the secondary windings by magnetic coupling. An additional switch $S_c$ connected to the secondary windings blocks the current in the load circuit during charging. 

\ctikzset{bipoles/length=1.4cm}

With respect to pulse transformer, several researcher applied superconducting coils as primary transformer windings and simultaneoulsy as opening switches \cite{Hilal1987,Singh1988,Floch1999}. The current in the primary windings was rapidly reduced by forcing the superconductor into the normal-conducting state by an intended temperature increase. Amplification factors of more than 20 were reported. Nonetheless, the use of a pulse transformer as power supply for a railgun system was first reported by the \gls{TTU}. Their railgun facility HERA comprised a pulse transformer which was connected to a 500\,kJ capacitive \gls{PFN} which can deliver a maximum current of 100\,kA \cite{Woffort1991,Day1993}. The turns ratio of the transformer is at maximum 20:1, but it is operated with a turns ratio of 5:1. Pulse compression is not applied in this system. Another system was very recently built at the GEDI Lab for the supply of a brush railgun \cite{Dreizin2015}. Here, a modular 14\,MJ double layer capacitor bank is supplying a pulse transformer with a maximum current of 70\,kA which is then amplified by the transformer with a winding ratio of 20:1. The transfer to the load and pulse compression is initiated by a pneumatically driven opening switch. Higher amplification factors can be expected by combining pulse transformer circuits in an XRAM-like topology as done in \cite{Li2012}. Here a transformer with superconducting primary windings and \acrshort{IGBT} switches were used. Single stage experiments show that a seed current of 100\,A was amplified to 4.7\,kA \cite{Li2012b}. Based on this small scale experiment and simulation results, it was estimated that several stages can amplify a current of 500\,A to more than 120\,kA. 

\subsection{Meat Grinder}
\begin{figure*}[!ht]
  \footnotesize
  \centering
  \input{Figs/Inductive/MEATgrindercircuit.tex}
  \caption{Basic meat grinder circuit}
  \label{fig:Meatgrinder}
\end{figure*}
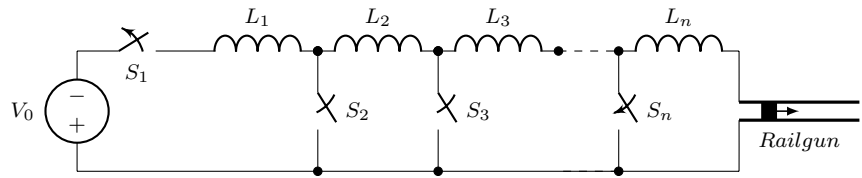

In the meat grinder concept, several coupled coils are charged in series by a common current source (see Fig.~\ref{fig:Meatgrinder}). Opening switches are used to subsequently disconnect the coils from the circuit until one coil is left which is connected to the load. In the first step, $S_1$ is opened and $S_2$ is closed simultaneosly. The switches of the following stages are switched likewise in subsequent steps ($S_2$ opens when $S_3$ closes and so on). During each switching process, the energy of the disconnected coil is transferred to the remaining coils by magnetic field coupling. Thus the load current can be increased or decreased \cite{Zucker1983,Kratz2002}. In fact, the meat grinder can be seen as a special type of pulse transformer where primary and secondary windings are galvanically connected instead of being insulated.

%
\ctikzset{bipoles/length=1.4cm}

The meat grinder has been under consideration as railgun power source for more than 30 years. 
In this respect, a single step meat grinder was tested with a 1\,m-long railgun as load by Lindner et al. \cite{Lindner1987}. Experiments with primary energies of up to 42\,kJ and current amplitudes of 150\,kA were conducted but did not meet the predicted current amplification as the fuse opening switch did not open in time. Twenty years later, Sitzman et al. improved the concept by adding a storage capacitor and called it the \acrshort{STRETCH} meat grinder \cite{Sitzman2005}. They reported successful railgun experiments with a corresponding demonstrator which was realized by a multi-section Brooks coil, \acrshortpl{GTO} as opening switches and batteries as primary energy source \cite{Sitzman2007,Sitzman2007a}. It was claimed that with 30\,kJ of inductively stored energy, an output current amplitude of 172\,kA could be achieved \cite{McNab2014a,Sitzman2007a} but only preliminary experimental results with a current amplification from 2\,kA to 21\,kA were published. Recently, Wu et al. realized a meat grinder with a superconducting coil and \acrshortpl{IGBT} \cite{Wu2013,Wu2013b}. They amplified a charging current of 100\,A to 4.29\,kA at a resistive dummy load. X. Yu et al. are investigating the \acrshort{STRETCH} meat grinder with the \acrshort{ICCOS} switch as developed at the ISL \cite{Yu2013,Yu2013b,Yu2015a,Yu2015b,Yu2015c}. Most of their work is still conceptual but the increasing number of papers and involved personnel is indicating growing efforts.

\subsection{XRAM generator}
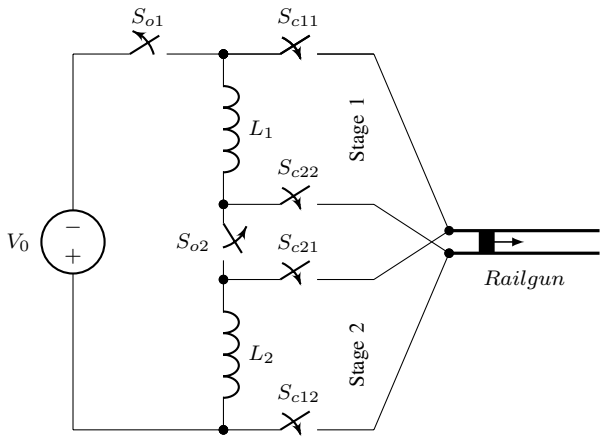
\begin{figure}[t]
  \footnotesize
  \centering
  \input{Figs/Inductive/XRAMcircuit.tex}
  \caption{Schematic for an XRAM generator with 2 stages}
  \label{fig:XRAMcircuit}
\end{figure}

In the XRAM circuit, inductors energized in series from a current source (battery, rotating machines, or others) are switched by corresponding opening and closing switches to discharge in parallel, producing an output current which is the sum of the individual inductor currents. In Fig.~\ref{fig:XRAMcircuit}, the source current is amplyfied by 2 after switching all switches synchronously. A higher amplification is possible with more than 2 stages. In high-energy pulsed-power systems, this multiplier technique simplifies the primary power design at the expense of more complex switching. The term XRAM generator originates from the well-known Marx generator where capacitors are charged in series and discharged in parallel to generate high voltages. By reversing the letters, the term XRAM indicates that the generator is the inductive counterpart to the Marx generator and its goal is to generate high currents. First proposed by Koch in 1967 \cite{Koch1967}, this topology was investigated several times as a power source for railguns \cite{Salge1985,Loeffler1988,Ford1993
,Kanter1993,Weck1997,Stallings2001
}. However, the largest devices of such kind were built in russia for other applications: a 12.5\,MJ-system within the X-ray test facility BAIKAL at TRINITI \cite{Grabovsky2001}, and an XRAM system with a 20\,MJ inductive storage at ESRI (see Tab.~\ref{tab:CoilGen}). 
Currently, XRAM generators for railguns are under investigation in Japan, China and France \cite{Yamada2014,Yu2013,Liebfried2013,Badel2012}. 
The XRAM concept was already proven to be applicable to superconducting coils \cite{Weck1997,Badel2010,Dedie2011b}. It is also promising together with an augmented railgun, then also known as XRAM-gun \cite{Salge1985,Badel2012,Liebfried2014}.

\ctikzset{bipoles/length=1.4cm}
\section{Superconducting magnetic energy storage}
Ohmic losses can be avoided by using superconducting coils. In this case, the current could be stored for very long time periods. Instead of electrical losses, thermal losses of the cooling system have to be considered. The maximum energy and power of a superconducting coil is limited by the maximum breakdown voltage, the operating conditions of the conductor material and its mechanical strength \cite{Badel2010}. Both the mechanical support and the electrical isolation will increase the coil volume and mass. Additionally, the cooling system has to be considerred. Therefore, superconducting coils for high-power applications are still very bulky. 
Nowadays, the realized \gls{SMES} systems in the order of 1\,MJ/1\,MW have a volume exceeding 2\,m$^3$ (estimated from \cite{Mito2009}). When the corresponding vacuum vessel, vacuum pumps and cooling equipment are taken into account, the volume can easily be multiplied by a factor of two or more. 
Therefore, a normal conducting coil seems presently more appropriate if a high energy density is desired.

However, several researchers applied superconducting coils in small scale demonstrators \cite{Singh1988,Li2012b,Badel2012,Wu2013}. Some of them applied the superconductor itself as opening switch by letting it quench on purpose \cite{Floch1999,Hilal1987,Weck1999}. A 20\,MJ/1\,MA conceptional design for a high temperature superconducting coil pulsed power generator is presented in \cite{Aso2011,Yamada2012}.

\section{Conclusion}
Inductive storage systems are applied to nearly all pulsed power generators, in some cases for voltage generation (HPGs, batteries) and in other cases for pulse shaping (capacitors, FCGs). They are theoretically more compact than capacitors but less compact than flywheels. As they need to be operated with a primary energy storage which can deliver a high current to the coil, the complete system compactness has to be considered. Flywheels, batteries and double-layer capacitors are available options in this respect. The latest research tends toward batteries as there are huge advances in the field of battery technology due to a large interest from the automotive industry. 

A challenge regarding inductive storages are still the opening switches for currents in the range of megaamperes. Recent developments show that  opening switches based on semiconducting switches will be possible but were not yet tested at an energy and power which is required for large railgun systems. Several inductive circuit topologies like the meat grinder or XRAM topologies are possible means of reducing the stress on opening switches by current amplifying properties. Table~\ref{tab:CoilGen} summerizes some of the most considerable inductive generators which were developed with respect to electromagnetic launch.

\input{TableInductors}

Generators based on inductive storage are an attractive alternative to capacitors, due to their higher energy density. This can be an asset when mobile platforms are concerned but not for stationary applications. It is especially interesting when the inductive storage can be integrated into the railgun itself, augmenting the driving magnetic field and thus improving the energy efficiency. If the railgun system is installed on an electromobile platform, one can expect that an energy storage system such as a battery or a flywheel is already available. In this case, an inductive pulsed power generator is the most convenient choice as it does not require an additional energy converter. 
The compactness of an inductive pulsed power generator should be even more pronounced compared to capacitors if bursts of repetitive discharges are envisaged.  

\bibliographystyle{IEEEtran}
\bibliography{IEEEabrv,PaperInductorReview}
%



%
\begin{IEEEbiography}[{\includegraphics[width=1in,height=1.25in,clip,keepaspectratio]{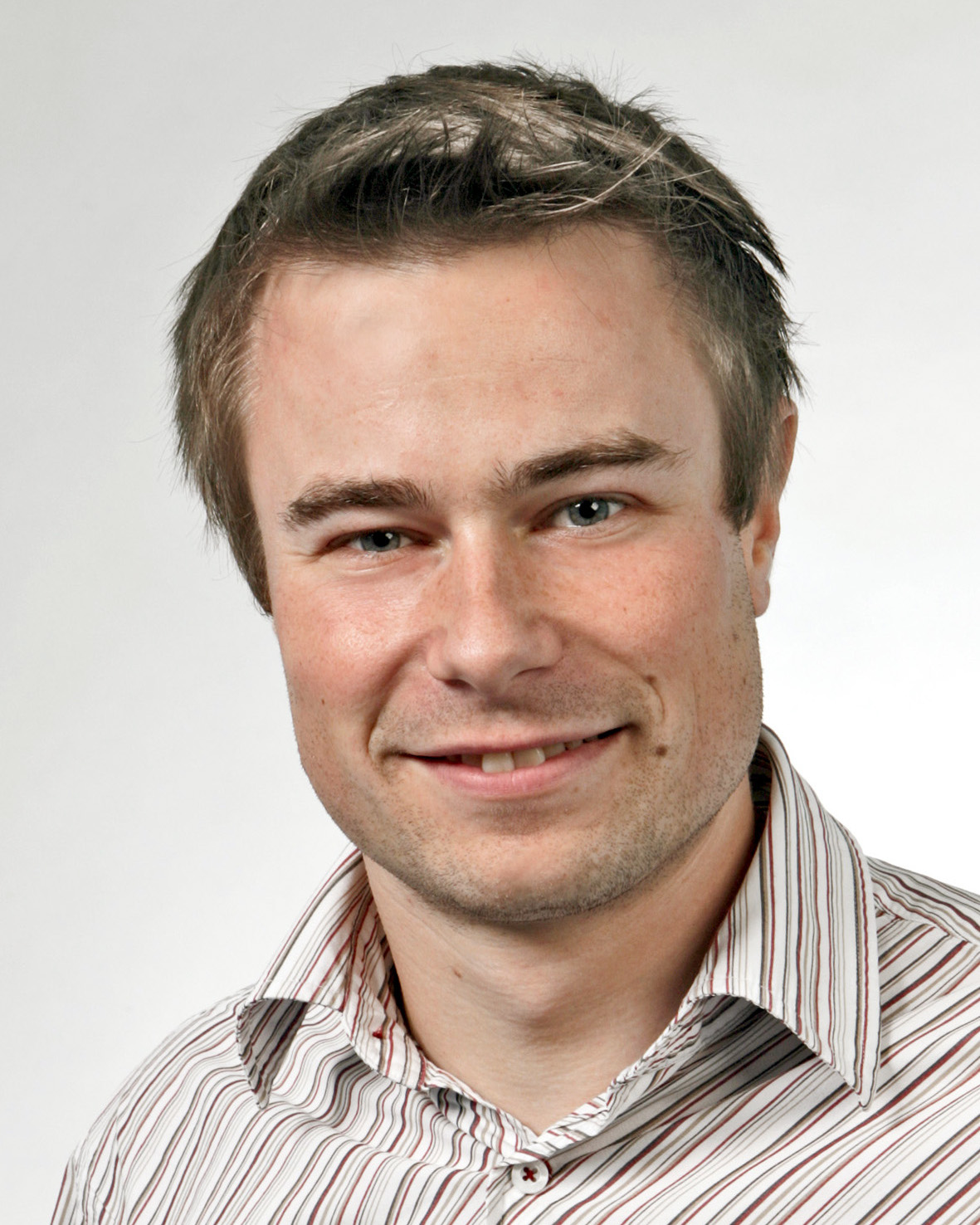}}]{Oliver Liebfried}
was born in 1980 in Bottrop, Germany. He received the Dipl.Ing.(FH) degree in electrical engineering in 2004 and the M.Eng. degree in energy system technology in 2006 from the Gelsenkirchen University of Applied Sciences, Gelsenkirchen, Germany. He received his Ph.D. degree from the Vilnius Gediminas Technical University in Vilnius, Lithuania. As Ph.D. student he was with the State Research Institute Center for Physical Sciences and Technology, Vilnius, Lithuania, and the French-German Research Institute of Saint-Louis, France, where he is currently working as research scientist. His research interests are capacitive and inductive pulsed power generators, electromagnetic accelerators, battery technology and metrology.
Dr. Liebfried is an executive board member of the International Society on Pulsed Power Applications e.V. (www.pulsed-power.org).
\end{IEEEbiography}






\vfill


\end{document}

%% file: PPgloss.tex

\newacronym[longplural={pulse power supplies}]{PPS}{PPS}{pulse power supply}
\newacronym[longplural={compact pulse power supplies}]{CPPS}{CPPS}{compact pulse power supply}
\newacronym{AC}{AC}{alternating current}
\newacronym{DC}{DC}{direct current}
\newacronym{SMES}{SMES}{superconducting magnetic energy storage}
\newacronym{G}{Gen}{Generator}
\newacronym{IES}{IES}{intermediate energy storage}
\newacronym{PES}{PES}{pulse energy storage}
\newacronym{PS}{PS}{pulse shaper}
\newacronym{RG}{RG}{railgun}
\newacronym{CS}{CS}{closing switch}
\newacronym{OS}{OS}{opening switch}
\newacronym{FCG}{FCG}{flux compression generator}
\newacronym{MFCG}{MFCG}{magnetic flux compression generator}
\newacronym{HPG}{HPG}{homopolar generator}
\newacronym[longplural={batteries}]{Batt}{Batt}{Battery}
\newacronym{SCap}{SCap}{supercapacitor}
\newacronym{PA}{PA}{pulsed alternator}
\newacronym{PVDF}{PVDF}{polyvinylidenflourid}
\newacronym{PFU}{PFU}{pulse forming unit}
\newacronym[longplural={distributed energy store}]{DES}{DES}{distributed energy store}
\newacronym{RAG}{RAG}{rotating arc gap}
\newacronym{GPW}{GPW}{given piece of wire}
\newacronym{GTO}{GTO}{gate turn-off thyristor}
\newacronym{IGCT}{IGCT}{integrated gate commutated thyristor}
\newacronym{IGBT}{IGBT}{insulated gate bipolar transistor}
\newacronym{STRETCH}{STRETCH}{slow transfer of energy through capacitive hybrid}
\newacronym{PFN}{PFN}{pulse forming network}
\newacronym{SCR}{SCR}{silicon controlled rectifier}
\newacronym{SC}{SC}{superconductor}
\newacronym{DED}{DED}{delivered energy density}
\newacronym{CPA}{CPA}{compensated pulsed alternator}
\newacronym{FTP}{FTP}{focussed technology programm}
\newacronym{Liion}{Li-ion}{lithium-ion}
\newacronym{LTO}{LTO}{lithium-titanate}
\newacronym{RnD}{R\&D}{research and development}
\newacronym{LTT}{LTT}{light triggered thyristor}
\newacronym{ICCOS}{ICCOS}{inverse current commutation with semiconductor devices}
\newacronym{MOSFET}{MOSFET}{metal oxide semiconductor field effect transistor}
\newacronym{RLC}{RLC}{resistor-inductor-capacitor (oscillation circuit)}

\newacronym{SiC}{SiC}{siliciumcarbid}
\newacronym{GaN}{GaN}{galiumnitrid}
\newacronym{NaK}{NaK}{sodium potassium}
\newacronym{NiMH}{NiMH}{nickel-metal-hydride}
\newacronym{LiCoO2}{LiCoO$_2$}{lithium-cobalt-oxide}
\newacronym{LiS}{Li/S}{lithium-sulfur}
\newacronym{LiMn2O4}{LiMn$_2$O$_4$}{lithium-ion-manganese-oxide}
\newacronym{LiO2}{LiO$_2$}{lithium-air}
\newacronym{LiFePO4}{LiFePO$_4$}{lithium-iron-phosphate}

\newacronym{ISL}{ISL}{French-German Research Institute of Saint-Louis}
\newacronym{TTU}{TTU}{Texas Tech University}
\newacronym{EARC}{EARC}{Electric Armaments Center}
\newacronym{IAT}{IAT}{Institute of Advanced Technologies}
\newacronym{KERI}{KERI}{Korea Electrotechnology Research Institute}
\newacronym{DERA}{DERA}{Defence Evaluation and Research Agency}
\newacronym{TZN}{TZN}{Technologiezentrum  Nord Forschungs- und Entwicklungszentrum Unterl\"u\ss~GmbH}
\newacronym{ARL}{ARL}{U.S. Army Research Laboratories}
\newacronym{NRL}{NRL}{U.S. Naval Research Laboratories}
\newacronym{NSWCDD}{NSWCDD}{U.S. Naval Surface Warfare Center Dahlgren Division}
\newacronym{HUST}{HUST}{Huazhong University of Science and Technology}
\newacronym{BISET}{BISET}{Bejing Institute of Special Electromechanical Technology}
\newacronym{IEE}{IEE}{Institute of Electrical Engineering}
\newacronym{CAS}{CAS}{Chinese Academy of Science}
\newacronym{SJU}{SJU}{Southwest Jiaotong University}
\newacronym{ECR}{ECR}{Energy Compression Research Corporation}
\newacronym{NIFS}{NIFS}{National Institute for Fusion Science}
\newacronym{JSW}{JSW}{Japan Steel Works, Ltd.}
\newacronym{IGE}{IGE}{Energy Engineering Institute}
\newacronym{CREEBEL}{CREEBEL}{Belfort Research Center in Electrical Engineering}
\newacronym{ESRI}{ESRI}{Efremov Scientific Research Institute}
\newacronym{ELR}{ELR}{Electromagnetic Launch Research, Inc.}
\newacronym{SNRC}{SNRC}{Soreq Nuclear Research Center}
\newacronym{UT}{UT}{The University of Texas at Austin}
\newacronym{CEM}{CEM}{Center for Electromechanics}
\newacronym{TNO}{TNO}{Netherlands Organisation for Applied Scientific Research - Pulse Physics Laboratory}
\newacronym{RARDE}{RARDE}{Royal Armament Research and Development Establishment}
\newacronym{IRD}{IRD}{Int. Research and Development, Ltd.}
\newacronym{WMD}{WMD}{Westinghouse Marine Division}
\newacronym{USAF}{USAF}{U.S. Air Force}
\newacronym{ARDEC}{ARDEC}{U.S. Army Armament Research, Development and Engineering Center}
\newacronym{MAI}{MAI}{Moscow Aviation Institute}

\newacronym{US}{U.S.}{United States}
\newacronym{USA}{USA}{United States of America}
\newacronym{NL}{NL}{Netherlands}
\newacronym{GB}{GB}{Great Britain}
\newacronym{UK}{UK}{United Kingdom}
\newacronym{USSR}{USSR}{Union of Soviet Socialist Republics}
\newacronym{RU}{RU}{Russia}
\newacronym{FR}{FR}{France}
\newacronym{DE}{DE}{Germany}
\newacronym{KR}{KR}{South Korea}
\newacronym{CN}{CN}{China}

\newacronym{NA}{N/A}{not available/not applicable}

%% file: Figs/Inductive/Singlecoilcircuit.tex

  \begin{circuitikz}[scale=1,american]
    \draw
    (0,0) to [american voltage source,l^=$V_0$](0,2) to [opening switch,l_=$S_o$,-*](2,2) to [L,l^=$L_{store}$,*-*](2,0) to [short,i_=$I_{charge}$] (0,0)
    (2,0) to [short,i=$I_{load}$](4,0)
    (4,2) to [switch,l^=$S_c$,-*](2,2)
    (4,2) to [short] (4,1.15) 
    (4,0.85) to [short] (4,0);
    \draw[very thick] (4,1.15) to [short] (6,1.15)
    (6,0.85) to [short] (4,0.85);
    \draw[line width=6pt] (4.5,1.15) -- (4.5,0.85);
    \draw[-latex,semithick] (4.5,1) -- (5,1);
    \draw (5,0.5) node {$Railgun$};
 \end{circuitikz} 

%% file: Figs/Inductive/Transformercircuit.tex

  \begin{circuitikz}[scale=1,american]
    \draw
    (0,0) to [american voltage source,l^=$V_0$](0,2) to [opening switch,l_=$S_o$](2,2) to [L] (2,0) to [short] (0,0);
    \draw (3,1.42) to [short] (3,2) to [switch,l=$S_c$] (5,2) to [short] (5,1.15)
    (3,0.58) to [short] (3,0) to [short] (5,0) to [short] (5,0.85);
    \draw (3,1.42) [thick] arc (90:270:0.2cm and 0.14cm) (3,1.14) arc (90:270:0.2cm and 0.14cm) (3,0.86) arc (90:270:0.2cm and 0.14cm) (3,0.58);

    \draw[very thick] (5,1.15) to [short] (7,1.15)
    (7,0.85) to [short] (5,0.85);
    \draw[line width=6pt] (5.5,1.15) -- (5.5,0.85);
    \draw[-latex,semithick] (5.5,1) -- (6,1);
    \draw (6,0.5) node {$Railgun$};
 \end{circuitikz}

%% file: Figs/Inductive/MEATgrindercircuit.tex

  \begin{circuitikz}[scale=0.8,american]
    \draw
    (0,0) to [american voltage source,l^=$V_0$](0,2) to [opening switch,l_=$S_1$](2,2) to [L,l^=$L_{1}$,-*](4,2) to [closing switch,l^=$S_2$,*-*](4,0) to [short] (0,0);
    \draw (4,2) to [L,l^=$L_{2}$,*-*] (6,2) to [closing switch,l^=$S_3$,*-*](6,0) to [short] (4,0);
    \draw (6,2) to [L,l^=$L_{3}$,*-*] (8,2) 
    (9,2) to [ switch,l^=$S_n$,*-*](9,0) to [short] (6,0);
    \draw (9,2) to [L,l^=$L_{n}$] (11,2) to [short] (11,1.15)
    (11,0.85) to [short] (11,0) to [short] (9,0);
    \draw[dashed] (8,2) -- (9,2)
    (9,0)  -- (8,0);  
    
    \draw[very thick] (11,1.15) to [short] (13,1.15)
    (13,0.85) to [short] (11,0.85);
    \draw[line width=6pt] (11.5,1.15) -- (11.5,0.85);
    \draw[-latex,semithick] (11.5,1) -- (12,1);
    \draw (12,0.5) node {$Railgun$};
    
 \end{circuitikz} 

%% file: Figs/Inductive/XRAMcircuit.tex

  \begin{circuitikz}[scale=1,american]
    \draw(0,5) to [L,l^=$L_1$,*-*](0,3) to [opening switch,l_=$S_{o2}$](0,2) to [L,l^=$L_2$,*-*](0,0);
    \draw(0,0) to [short,*-](-2,0) to [american voltage source,l^=$V_{0}$](-2,5) to [opening switch,l^=$S_{o1}$,-*](0,5) to [short,*-*](0,5) to [switch,l^=$S_{c11}$,*-](2,5);
    \draw(0,0) to [switch,l^=$S_{c12}$,*-](2,0) to [short,-*](3,2.35)
     (0,5) to [switch](2,5) to [short] (3,2.65);
    \draw(0,3) to [switch,l^=$S_{c22}$,*-](2,3) to [short,-*](3,2.35);
    \draw(0,2) to [switch,l^=$S_{c21}$,*-] (2,2) to [short,-*](3,2.65);
    \draw (1.8,4) node[rotate=90] {Stage 1};
    \draw (1.8,1) node[rotate=90] {Stage 2};

    \draw[very thick] (3,2.65) to [short] (5,2.65)
    (5,2.35) to [short] (3,2.35);
    \draw[line width=6pt] (3.5,2.65) -- (3.5,2.35);
    \draw[-latex,semithick] (3.5,2.5) -- (4,2.5);
    \draw (4,2) node {$Railgun$};
    
  \end{circuitikz} 

%% file: TableInductors.tex

\begin{sidewaystable*}[p]
\caption{Examples of realized pulsed power generators applying inductive storages}
	\label{tab:CoilGen}
	\centering
		\begin{tabular}[h]{|c|c|c|c|c|c|c|c|c|}
		\hline
			 Feeding &  &  Inductively &  & Max. load &  &  &  & \\
			 power supply & Switches &  stored energy & Load & current & Size & Operator & Year & Ref.\\
		\hline
	 \multicolumn{9}{|l|}{\cellcolor[gray]{0.9}Single coil topology}\\
		\hline
 			Battery & Fuse & 1.5\,MJ & Resistor & 28.5\,kA & 0.4\,m$^3$ & \acrshort{ELR}, \acrshort{UK} & 1985 & \cite{Gredin1985} \\ 	
  		\acrshort{HPG} & N/A & 3.1\,MJ & Railgun & 1000\,kA & 1\,m$^3$ $^a$ & \acrshort{CEM}-\acrshort{UT}, \acrshort{USA} & 1985 & \cite{Driga1985}\\
			\acrshort{HPG} & Exp. switch & 10\,MJ & Railgun & 2000\,kA & 3\,m$^3$ $^a$ & Westinghouse, \acrshort{USA} & 1986 & \cite{Hackworth1986}\\
			Capacitors & Thyristor/Diodes & 40\,kJ & Railgun & 52\,kA & N/A & \acrshort{ISL}, \acrshort{FR} & 1995 & \cite{Spahn1995}\\
			Capacitors & Ignitrons & 0.7\,MJ & Railgun & 250\,kA & 0.5\,m$^3$ & \acrshort{IAT}, \acrshort{USA} & 1995 & \cite{Parker1997}\\
			Capacitors & Thyristor/Diode & 60\,kJ & n/a & 68\,kA & \textgreater\,6\,dm$^3$ & \acrshort{ESRI}, \acrshort{RU} & 2009 & \cite{Kovrizhnykh2009}\\
		\hline
			\multicolumn{9}{|l|}{\cellcolor[gray]{0.9}Pulse transformer topology}\\
		\hline
		 	Battery & \acrshort{SCR}/\acrshort{GTO} & 16\,kJ & short-circuit & 150\,kA & N/A & \acrshort{TNO}, \acrshort{NL} & 1989 & \cite{Tuinman1989}\\		
			\acrshort{PFN} & Ignitrons/Diode &  550\,kJ $^\star$ & Railgun & 500\,kA & 1\,m$^3$  & \acrshort{TTU}/\acrshort{USA} & 1993 & \cite{Woffort1991,Day1993}\\
		 	\acrshort{DC} supply & Quenched \acrshort{SC} &  500\,J & N/A & 23.3\,kA & 3\,m$^{3 \hspace{0.1cm} a}$ & \acrshort{IGE}/\acrshort{CREEBEL}, \acrshort{FR} & 1999 & \cite{Floch1999}\\			
			\acrshort{DC} supply & \acrshort{IGBT}/Diode & 69\,J & Resistor & 4.26\,kA & 12\,dm$^3$$^{\ast}$ & \acrshort{SJU}, \acrshort{CN} & 2012 & 		\cite{Li2012b}\\
		\hline
			\multicolumn{9}{|l|}{\cellcolor[gray]{0.9}Meat grinder topology}\\
		\hline
			Capacitors & Fuse/Varistor &  19.1\,kJ & Railgun & 135\,kA & N/A  & \acrshort{ECR}, \acrshort{USA} & 1987 & \cite{Lindner1987}\\
			\acrshort{Liion} batteries & \acrshort{GTO}s &  30\,kJ & Railgun & 172\,kA & $\sim \frac{1}{8}$\,m$^3$$^{\ast}$  & \acrshort{IAT}, \acrshort{USA} & 2007 & \cite{Sitzman2007,Sitzman2007a,McNab2014a}\\
			\acrshort{DC} supply & \acrshort{IGBT} &  78\,J & Resistor & 4.29\,kA & \~0.003\,m$^3$  & \acrshort{SJU}, \acrshort{CN} & 2013 & \cite{Wu2013,Wu2013b}\\
			\hline
			\multicolumn{9}{|l|}{\cellcolor[gray]{0.9}XRAM topology}\\
		\hline
			\acrshort{HPG} & Exp. switch & 20\,MJ & N/A & 6000\,kA & 316\,m$^3$ & \acrshort{ESRI}, \acrshort{RU} & 1977-1992 & \cite{Druzhinin1992}\\
			Generator & Exp. switches & 12.5\,MJ & X-ray system & 1500\,kA & N/A & TRINITI, RU & 2001 & \cite{Grabovsky2001}\\
			Battery & \acrshort{GTO}/Varistor &  25\,kJ & Resistance & 10\,kA & N/A & \acrshort{SNRC}, Israel & 1993 & \cite{Kanter1991,Kanter1993}\\	
			Capacitors & \acrshort{ICCOS} &  4725\,J & Cable & 60\,kA & 0.2\,m$^{3}$  & \acrshort{ISL}, \acrshort{FR} & 2011 & \cite{Dedie2011}\\
		 	Capacitors & \acrshort{ICCOS} &  200\,kJ & Railgun & 40\,kA & 0.4\,m$^{3}$  & \acrshort{ISL}, \acrshort{FR} & 2013 & \cite{Liebfried2013}\\
		 	Capacitors & Spark gap/Diode &  93\,kJ & Aug. Railgun & 200\,kA & integrated  & \acrshort{ISL}, \acrshort{FR} & 2014 & \cite{Liebfried2014}	\\
			Supercaps & \acrshort{IGBT}s &  25\,kJ & Cable$^{\ast}$ & 24\,kA & \textless\,1\,m$^3$ & \acrshort{NIFS}/\acrshort{JSW}, Japan & 2014 & \cite{Yamada2014}\\

		\hline
		\multicolumn{9}{|l|}{\cellcolor[gray]{0.9}$^{\star}$ capacitively stored \hspace{0.5cm} $^{\ast}$ estimated  \hspace{0.5cm}  $^a$ just coil}\\
		\hline
		\multicolumn{9}{|l|}{ELR - \acrlong{ELR}; CEM-UT - \acrlong{CEM}, \acrlong{UT}; IAT - Institute of Advanced Tech.;}\\
		\multicolumn{9}{|l|}{ESRI - Efremov Scientific Research Institute; TNO - \acrlong{TNO};	TTU - Texas Tech Univ.;}\\
		\multicolumn{9}{|l|}{IGE - Energy Eng. Institute; CREEBEL - Belfort Research Center in Electrical Eng.; SJU - Southwest Jiaotong Univ.;}\\
		\multicolumn{9}{|l|}{ECR - Energy Compression Research Corp.; SNRC - \acrlong{SNRC}; NIFS - National Institute for Fusion Science; JSW - Japan Steel Works, Ltd.}\\
		\hline
\end{tabular}	
\end{sidewaystable*}